\documentclass[fleqn,twoside]{article}%
\usepackage{amsmath}
\usepackage{amsfonts}
\usepackage{amssymb}
\usepackage{graphicx}%
\def\be{\begin{equation}}
\def\ee{\end{equation}}
\def\bi{\bibitem}

\begin{document}

\title{Can particle creation phenomena replace dark energy?}
\author{Subhra Debnath and Abhik Kumar Sanyal }
\maketitle
\begin{center}
Dept. of Physics, Jangipur College, Murshidabad, \noindent
West Bengal, India - 742213. \\
\end{center}

\begin{abstract}
Particle creation at the expense of gravitational field might be sufficient to explain the cosmic evolution history, without the need of dark energy at all. This phenomena has been investigated in a recent work by Lima et-al \cite{l:s} assuming particle creation at the cost of gravitational energy in the late Universe. However, the model does not satisfy the WMAP constraint on matter-radiation equality \cite{g:r}. Here, we have suggested a model, in the same framework, which fits perfectly with SNIa data at low redshift as well as early Integrated Sachs-Wolfe effect on matter-radiation equality determined by WMAP at high redshift. Such a model, requires the presence of nearly $26\%$ primeval matter in the form of baryons and CDM.
\end{abstract}

\section{Introduction}

Recently released 7-year WMAP data \cite{w1},\cite{w2} has found no trace of deviation from the standard $\Lambda$CDM model. But the problem in connection with the cosmological constant remains unresolved. The vacuum energy density, as calculated by the field theorists, is some $10^{120}$ order of magnitude greater than the cosmological constant $\Lambda$ required by the cosmologists, to explain late time cosmic acceleration, which is of the order of $H_{0}^2$, $H_0$ being the present Hubble parameter. So far, many alternatives to the standard $\Lambda$CDM model have been proposed and explored and as a matter of fact, all of these models have been found suitable to explain late time cosmic acceleration. The problems associated with these models are, - they can not be distinguished from the standard $\Lambda$CDM model at one hand, and most of them are not suitable to explain the early Universe, on the other. However, $\Lambda$CDM model requires $26\%$ of matter in the form of pressureless dust, out of which only $4\%$ are baryons and the rest, about $22\%$ are cold dark matter (CDM). Since dark matter interacts only with the gravitational field, so  it plays a key role in the structure formation. The gravitational Jeans instability allows compact structures to form and is not opposed by any force like radiation pressure in the case of dark matter. As a result, dark matter begins to collapse into a complex network of dark matter Halos, well before ordinary baryonic matter, which is impeded by pressure force. Without dark matter the epoch of Galaxy formation would have occurred at a substantially later stage, than observed. Thus, the amount of CDM ($22\%$) must have been created in the very early Universe, prior to the radiation dominated era together with the baryons, by some sort of mechanism, viz., supersymmetry breaking, cosmic string decay or particle creation at the expense of gravitational field. These particles are usually supposed to be weakly interacting massive particles (WIMP). For example, as a heavy stable particle, the lightest neutralino is an excellent candidate to comprise the Universe's cold dark matter. In many models (see \cite{br} for a nice review) the lightest neutralino can be produced in the hot early Universe and leave approximately the right relic abundance to account for the observed dark matter, ie., $22\%$ as required by $\Lambda$CDM model. Now, if phenomenologically one considers that CDM may also be produced by gravitational particle creation mechanism, even at a very slow rate, during the late time evolution of the Universe, viz., during the matter dominated era, as considered by Lima et-al \cite{l:s}, then it may be possible to explain the presently observable acceleration of the Universe, without taking dark energy into account. The very advantage of the creation of cold dark matter over dark energy is that it avoids coincidence problem and also may be detectable in future experiments. In the present work our focus is on the cosmological consequences of particle production on the evolution of late stage of the Universe which was initiated recently by Alcaniz and Lima \cite{j} and Lima, Silva and Santos(lss) \cite{l:s}. \\

\noindent We remember that in the eighty's, the motivation that initiated to go after inflationary scenario, was the fundamental three problems associated with Friedmann model, viz., the horizon, the flatness and the observed isotropy and homogeneity in the cosmic microwave background radiation (CMBR). There was an additional problem in the form of huge entropy per baryon in the observable Universe $(\sim 10^{87})$. Einstein's equations are purely adiabatic and reversible. Consequently, these equations can hardly provide, by themselves, an explanation relating to the origin of cosmological entropy. This problem was resolved \cite{2} by taking into account the cosmological consequence of irreversible particle creation phenomena in the framework of Einstein's equation, classically.\\

\noindent Particle creation phenomena was explored largely during the last century to explain the early Universe. Matter constituents may be produced quantum mechanically \cite{p}, \cite{bd}, \cite{mw} in the framework of Einstein's equations. Cosmological consequence of particle creation mechanism is studied taking into account an explicit phenomenological balance law (see appendix) for the particle number \cite{2}, \cite{3}, \cite{l} in addition to the familiar Einstein's equations. In view of such a balance law, Prigogine et-al \cite{2} successfully explained the cosmological evolution of the early Universe.  \\

\noindent Recently, Lima et-al \cite{l:s} have developed a late time model Universe, taking into account the creation phenomena in the matter dominated era. The model admits early deceleration followed by a recent acceleration of the Universe as suggested by present observations and fits SNIa data to some extent. However, they \cite{l:s} have shown that in their model the creation phenomena is never ending, as a result cosmic evolution does not ever include standard radiation dominated or matter dominated Friedmann era. This definitely creates problem in explaining the structure formation of the Universe and the CMBR. Later, Steigman et-al \cite{g:r} analyzed the model in the two limits of high and low redshifts. They have observed a clear conflict between the WMAP constraint on matter-radiation equality $z_{eq}$ at high redshift and SNIa data at low redshift. The main criticisms of the $\beta-\gamma$ model proposed by Lima et-al \cite{l:s} are that they have not taken into account the amount of CDM created in the very early Universe at one end, and that their creation rate $\Gamma = 3\beta H + 3\gamma H_0 $ depends on the present Hubble parameter, on the other. If these problems are alleviated, then phenomenologically particle creation process obviously unifies early inflation with late stage of cosmic acceleration in an elegant fashion. \\

\noindent In the present work, we propose a model where, instead of choosing the creation parameter $\Gamma$ arbitrarily, we have considered the experimentally verified fact that the Universe has recently entered an accelerated phase of expansion. As a result we have chosen the scale factor judiciously, such that particle creation could start again in the matter dominated era. This naturally alleviates the said problems and unifies the early and the late stages of cosmic evolution, in an elegant fashion. Additionally, the model fits perfectly with the WMAP constraint on matter-radiation equality $z_{eq}$ only if one considers the presence of nearly $26\%$ of primeval matter in the form of baryons and CDM. In view of such a model, particle creation phenomena is now able to explain the history of cosmic evolution from the very early Universe till date, without requiring dark energy at any stage and thus avoiding coincidence problem.\\

\noindent In the following section we write down the field equations incorporating the phenomena of particle creation, which is apparent through the balance equation. In section 3, we show how the conflict between high redshift and the low redshift data may be reduced in the model presented by Lima et al \cite{l:s}, just by accounting for some amount of CDM produced in the very early Universe. However, we also mention some more problems which are still associated with their \cite{l:s} model. In section 4, instead of choosing a form of the creation rate arbitrarily, we rather choose a form of the scale factor, suitable for a transition from the early deceleration to the late time cosmic acceleration, to understand the associated problems. This gives us insight to find a form of the creation rate $\Gamma$ associated with the so called intermediate inflation \cite{if}. Such a form of $\Gamma$ appears to be much elegant to study the cosmological evolution alleviating all the problems discussed. This has been done in section 5. Finally we end up with the conclusion in section 6 and an appendix in section (7), to calculate the balance law.\\

\section {Balance law and the Field equations}

As mentioned in the introduction, cosmological consequence taking into account particle creation phenomena is studied by using an explicit phenomenological balance law for the particle number \cite{2},\cite{3}. Such a balance law in the process of particle production is modeled by $(nu^{\alpha})_{;\alpha}=\Psi,$ with a source term $\Psi$. For a vanishing $\Psi$ the particle number is conserved. In the isotropic and homogeneous background metric,

\be ds^2=-dt^2+a(t)^2\left[\frac{dr^2}{1-kr^2}+r^2 (d\theta^2+sin^2\theta d\phi^2)\right] \ee

\noindent above balance law reduces to, $\frac{\dot N}{N} = \frac{\dot n}{n}+\Theta = \frac{\Psi}{n} = \Gamma,$ where, $u^{\alpha}$ , $N$ , $n$ and $H$ are the fluid four velocity vector, the total number of particles, particle number density and the Hubble parameter respectively, while, $\Theta = u^{\alpha}\;_{;\alpha} = 3 H$ is the expansion scalar.
Thus, the field equations in the spatially flat ($k = 0$) Robertson Walker metric (1) associated with particle creation phenomena are,

\be 2\dot H + 3H^2 = - 8\pi G(p_m+p_{cm}),\ee

\be 3H^2  =  8\pi G (\rho_m + \rho_{cm}) = 8\pi G \rho,\ee

\be 3H + \frac{\dot n}{n} = \Gamma = \frac{\Psi}{n},\ee

\be p_{cm} = -\frac{\rho + p_m}{3H}\Gamma.\ee
\noindent
For the last equation (5) please see the appendix. In the above set of equations, $H = \frac{\dot a}{a}$ is the Hubble parameter, $p_m$ and $\rho_m$ are the pressure and the energy density of the matter existing in the Universe in the form of a barotropic fluid containing baryons and cold dark matter, created in the very early Universe. $p_{cm}$ and $\rho_{cm}$ are the pressure and the energy density of the cold dark matter in the form of WIMP created at the late stage of cosmic evolution, ie., during matter dominated era. $\Gamma$ is the creation parameter, and $n$ is the particle number density. It has been shown in the appendix that the second law of Thermodynamics allows the creation of particle from the gravitational field and the process is irreversible. Thus the creation parameter $\Gamma > 0$, and so it is clear from equation (5) that the particle creation phenomena is always associated with a negative pressure $p_{cm}$, which may be responsible for acceleration at the late stage of cosmic evolution. We would like to mention at this stage that, other than cold dark matter and baryons, different types of particles may be created in view of quantum field theory in curved space time \cite{p}, \cite{bd}, \cite{mw} and all are associated with a negative pressure, as mentioned. However, also, as mentioned in the introduction that the structure formation requires $22\%$ that are constituted by dark matter, which must have been created in the early Universe. The present estimated amount of baryons is $4\%$ and the rest amount required for present cosmic acceleration is usually treated as dark energy. Other form of matter (like hot dark matter, say) has negligible contribution. Here, we proceed to show that creation of the same amount of CDM also solves the puzzle. Further, note that the creation of baryons and CDM in the early Universe was also associated with a large negative pressure. However, the creation phenomena weakened and finally stopped, when Universe expanded sufficiently, thereby giving way to the hot big bang followed by the radiation dominated era of Friedmann type ($a \propto t^{\frac{1}{2}}$) \cite{2}. There after, the baryons and the CDM created in the very early Universe obviously start acting as pressureless dust, so that we can take $p_m = 0$ and hence $\rho_m = \rho_0 a^{-3}$, $\rho_0$ being a constant. Thus we can simplify the above equations to get,

\be \Gamma = 3 H + 2\frac{\dot H}{H} = 3H\left(\frac{2\dot H+3H^2}{3H^2}\right) = -3Hw_{e},\ee

\be \rho_{cm} = \frac{3H^2}{8\pi G} - \rho_0 a^{-3},\ee

\be p_{cm} = - \frac{1}{8\pi G}H\Gamma,\ee

\noindent $w_e$ being the effective state parameter. In view of the above set of three equations (6) - (8), we need to
find the scale factor $(a)$ (and consequently $(H)$, the Hubble parameter), the creation rate $\Gamma$, the creation pressure $p_{cm}$ and the creation matter density $\rho_{cm}$. Obviously, we need yet another suitable condition to solve the system of equations. Lima et al \cite{l:s} studied these equations under the assumption of a form of the creation rate $\Gamma$. Afterwards, while further studying their model in connection with data fitting, they found a clear conflict between the low (SNIa) and high (WMAP constraint on $z_{eq}$) \cite{g:r} redshift limits. In the
following section, we review the problem and show how the production of CDM in the very early Universe alleviates the problem.

\section{A brief review of lss model}

\noindent
The Friedmann equation, taking into account the created matter, baryonic matter and radiation reads,

\be \left(\frac{H}{H_0}\right)^2 = \Omega_r (1+z)^4 + \Omega_B (1+z)^3 + \frac{\rho_{cm}}{\rho_c},\ee

\noindent
where, $\rho_c$ is the present value of critical density. To calculate the last term let us take the total number of created particles at an instant to be $N = nV$, where $V = V_0 (1+z)^{-3}$ is the comoving volume at that instant.
Thus the creation rate is given as,

\be \frac{1}{N}\frac{dN}{dt} = \frac{d[\ln{(\rho_{cm}V)}]}{dt} = \Gamma,\ee

\noindent
which yields,

\be \rho_{cm} = \rho_{cm0}(1+z)^3 \exp{(-\int_{t}^{t_0}\Gamma dt')},\ee

\noindent
where, $\rho_{cm0}$ is the present value of the created matter density. Thus,

\be \frac{\rho_{cm}}{\rho_c} = \Omega_{cm}(1+z)^3 \exp{(-\int_{t}^{t_0}\Gamma dt')},\ee

\noindent
where, $\Omega_{cm}$ is the density parameter corresponding to the created matter. So, the Friedmann equation (9) finally reads,

\be \left(\frac{H}{H_0}\right)^2 = \Omega_r (1+z)^4 + \Omega_B (1+z)^3 + \Omega_{cm}(1+z)^3 \exp{(-\int_{t}^{t_0}\Gamma dt')}.\ee

\noindent
Now, under the assumption $\Gamma = 3\beta H + 3\gamma H_0$, where, $\beta$ and $\gamma$ are constants and $H_0$ is the present Hubble parameter, (lss) \cite{l:s} obtained a solution of the scale factor in the form,

\[a(t) = a_0 \left[\frac{1-\gamma-\beta}{\gamma}\left(e^{\frac{3\gamma H_0 t}{2}}-1\right)\right]^{\frac{2}{3(1-\beta)}},\]

\noindent
which admits the observed transition from early deceleration to late time acceleration. In a later investigation \cite{g:r}, this model was found to produce a conflict between SNIa data at low redshift and WMAP - 5 year data constraint \cite{w3} on mater-radiation equality $z_{eq} = 3141 \pm 157$, occurred at the high redshift limit of observed ISW effect. Let us review the situation to find the real problem associated with the conflict. The Friedmann equation (13) in the model under consideration reads,

\be \left(\frac{H}{H_0}\right)^2 = \Omega_r (1+z)^4 + \Omega_B (1+z)^3
+ \Omega_{cm}(1+z)^{3(1-\beta)} \exp{3\gamma(\tau-\tau_0)},\ee

\noindent
where, $\tau = H_0 t$ and $\tau_0 = H_0 t_0$ are the age at any instant and the present age of the Universe respectively, in the units of Hubble's age $(H_0^{-1})$. Setting, $\Omega_{cm} = 1 - \Omega_B$, the $\gamma - \beta$ relation is obtained (see equation (34) in \cite{g:r}) as,

\be \gamma = (1-\beta)\left[(1-\Omega_{B})^{\frac{1}{2}}-\{\Omega_r(1+z_{eq})
-\Omega_B\}^{\frac{1}{2}}(1+z_{eq})^{\frac{3\beta}{2}}\right].\ee

\noindent
This model fits SNIa data for $\beta = 0$ and $\gamma = 0.66 \pm 0.04$, while $1 + z_{eq} = 1798^{+536}_{-552}$, taking $\Omega_{B} = .042$. Clearly, the model does not fit with the WMAP - 5 year data constraint \cite{w3} on mater-radiation equality $z_{eq} = 3141 \pm 157$, occurred at the high redshift limit of observed ISW effect. This contradiction may be alleviated easily, if we consider existence of cold dark matter that was created in the early Universe and which was responsible for inflation. As, already mentioned, this amount of CDM created in the very early Universe behaves now as pressureless dust and has been redshifted like baryons. If we now add corresponding density parameter $\Omega_{CDM}$, associated with the cold dark matter created in the very early Universe in equation (13), it reads

\be \left(\frac{H}{H_0}\right)^2 = \Omega_r (1+z)^4 + \Omega_m (1+z)^3 + \Omega_{cm}(1+z)^3 \exp{(-\int_{t}^{t_0}\Gamma dt')},\ee

\noindent
where, $\Omega_{m} =  \Omega_B + \Omega_{CDM}$ and, $\Omega_{cm} = 1 - \Omega_m$. In the absence of matter creation phenomena in the late Universe, $\beta, ~\gamma$ vanish, and hence $\Gamma = 0$. Thus there is no creation pressure $p_m$ as well as creation matter density $\rho_{cm}$. Hence, $\Omega_{cm} = 0$. Thus at the matter-radiation equality $(z_{eq} = \frac{\Omega_m}{\Omega_r}-1)$ taking, $\Omega_m = \Omega_B + \Omega_{CDM} = 0.26$ and $\Omega_r = 8\times 10^{-5}$, one recovers $z_{eq} = 3249$, which is at par with WMAP data. Equation (15) now takes the form,

\be \gamma = (1-\beta)\left[(1-\Omega_m)^{\frac{1}{2}}-\{\Omega_r(1+z_{eq})
-\Omega_m\}^{\frac{1}{2}}(1+z_{eq})^{\frac{3\beta}{2}}\right].\ee

\noindent
If we now consider that $16\%$ of CDM (say) were produced in the very early Universe, then $\Omega_m = \Omega_B + \Omega_{CDM} = 0.2$ and thus for $\beta = 0$ and $\gamma = 0.66 \pm 0.04$, $z_{eq} = 3186^{+254}_{-214}$, which is very much at par with WMAP data \cite{w1}, \cite{w2}, \cite{w3}. This clearly indicates that one should include the contribution of CDM created at the very early Universe. The creation of CDM in the very early Universe, as mentioned in the introduction, was halted and the Universe entered usual Friedmann radiation dominated era. Thereafter, this amount of CDM is being redshifted like baryons.\\

\noindent
In the above analysis, while we have showed how the conflict encountered between low and high redshift data \cite{g:r} may be reduced, nevertheless, it does not support the model \cite{l:s}. Firstly, in their $\beta -\gamma$ model, $\beta = 0$, somehow fits SNIa data, which is not a very good fit at all (see fig. 1 \cite{g:r}). Further, $\beta = 0$ turns out to give a constant creation rate throughout the evolution of the Universe, which is highly objectionable. Also, the choice of the creation parameter ($\Gamma$) as a function of present Hubble parameter ($H_0$) implies that the model is plagued by the coincidence problem. Finally, we could accommodate only $16\%$ of CDM out of $22\%$ to alleviate the conflict \cite{g:r}. Addition of another $6\%$ of CDM shifts $z_{eq}$ to a much higher value. In view of the above
criticism we pose to present a more realistic model.

\section{Case 1} To get an explicit solution of the field equations (6) through (8), we need yet another physically reasonable assumption. However, we really have no idea of the rate of matter creation $\Gamma$ either from quantum field theoretic (QFT in CST) or from classical kinetic approach. Nevertheless, it is clear from equation (5) that such phenomena is associated with a negative pressure $p_c$. If the creation pressure is sufficiently negative, it might lead to an accelerating phase of cosmological evolution. Additionally, in order to get an idea about the form of the creation rate we can also depend on the presently available cosmological data and the best fit models. Since $\Lambda$CDM model has excellent fit with the SNIa and WMAP data, so we can infer certain important aspects of cosmological evolution. First of these is definitely that the Universe has encountered a transition from early deceleration to late time acceleration. Next is, early growth of perturbation should track $\Lambda$CDM model closely. These facts allow us to choose a suitable form of the scale factor at par with the present experimental results and in the process, we expect to get an idea on the form of $\Gamma$. In view of the two aspects of the late time cosmological evolution just discussed, we choose the scale factor in the matter dominated era as,

\be a = a_0 t^{\frac{2}{3}}+b_{0}t^{\alpha} = b_0(q t^{\frac{2}{3}}+t^{\alpha}),\ee

\noindent
where, $a_0$, $b_0$ and ${\alpha} > 1$ are constants with, $q = \frac{a_0}{b_0} > 0$. With such a form of the scale factor, the first term of equation (18) plays the leading role in the early stage of cosmological evolution in the matter dominated era and the Universe tracks Friedmann model $a \propto t^{\frac{2}{3}}$, closely. Thus, the growth of perturbation in connection with the structure formation tracks $\Lambda$CDM model closely. The second term appears in connection with the creation of matter, which is associated with a negative creation pressure. As $t$ increases, second term starts playing significant role and it starts dominating as $t^{\alpha -\frac{2}{3}} > q$, which implies that the creation phenomena starts rather late. Finally, since $\ddot a = b_0[-\frac{2q}{9 t^{\frac{4}{3}}}+\alpha(\alpha - 1)t^{\alpha-2}]$, so acceleration ($\ddot a > 0$) starts, only when $t > \left[\frac{2q}{9\alpha(\alpha-1)}\right]^{\frac{3}{3\alpha -2}}$. Thus, equation (18) clearly depicts early deceleration and late time acceleration of the Universe. Now, in the matter dominated era, $p_m = 0$, the solutions in view of the chosen form of the scale factor (18) are,

\be H = \frac{\frac{2}{3}q t^{-\frac{1}{3}}+{\alpha} t^{{\alpha}-1}}{q t^{\frac{2}{3}}+t^{\alpha}}.\ee

\be \Gamma = -3Hw_{e} = 2\frac{\dot H}{H}+3H = \frac{(3 {\alpha} - 2) \left(2 (3 {\alpha} + 1) q t^{\frac{2}{3}} + 9 {\alpha} t^{\alpha}\right)t^{({\alpha} - 1)}}{3 (q t^{\frac{2}{3}} + t^{\alpha}) (2 q t^{\frac{2}{3}} + 3 {\alpha} t^{\alpha})}.\ee

\be p_{cm} = -\frac{1}{8\pi G} \Gamma H = -\left[\frac{ (3 \alpha-2) \left(
    2 (3 \alpha + 1)q t^{\frac{2}{3}} + 9\alpha t^\alpha  \right)t^{(\alpha-2)}}{
 72 \pi G (q t^{\frac{2}{3}} + t^\alpha)^2}\right].\ee

\be \rho = \rho_{m}+\rho_{cm}  = \frac{3H^2}{8\pi G} = \frac{(2 q t^{\frac{2}{3}} +
  3\alpha t^\alpha )^2}{24 \pi G  t^2 (q t^{\frac{2}{3}} + t^\alpha)^2}. \ee

\be w_{e} = -\frac{2\dot H +3H^2}{3H^2} = -\left[\frac{(3 \alpha-2) \left(2(3 \alpha + 1)q t^{\frac{2}{3}} + 9 \alpha t^\alpha \right)t^{\alpha}}{3 (2 q t^{\frac{2}{3}} + 3 \alpha t^\alpha )^2}\right].\ee
\[w_{cm} = -\frac{2\dot H +3H^2}{3H^2 -8\pi G\rho_{m}} = -\frac{2\dot H +3H^2}{3H^2 - 3 H_0^2 \Omega_m (1+z)^3}.\]
\be z =\frac{ q t_0^{\frac{2}{3}}+ t_0^{\alpha}}{q t^{\frac{2}{3}}+t^{\alpha}}-1.\ee

\noindent
In the above set of solutions $w_e$ and $w_{cm}$ are the effective state parameter and the state parameter corresponding to the created matter respectively. Since $\alpha > 1$, so the expression (23) for $w_e$ is clearly negative which, as mentioned earlier, implies that the created matter is such that it is always associated with a negative creation pressure. \\

\noindent
\textbf{Fitting  the observational data.}\\

\noindent
The present model is parametrized by the two parameters ${\alpha}$ and $q$. With ${\alpha} = 4, ~h = \frac{9.78}{H_0^{-1}} = 0.7 ~Gyr^{-1}$, and fixing $H_0 t_0 = 1$, $q$ is automatically fixed and the distance modulus versus redshift curve (blue) is found to fit perfectly with the $\Lambda$CDM model (red). In fact the two are practically indistinguishable (fig.1). We get the present value of the  effective state parameter $w_{e0} = -1$ and the transition redshift $z_a = 0.56$, which are in excellent agreement with $\Lambda$CDM model. It is observed that the effective state parameter encounters a transient double crossing of the phantom divide line in the future (fig.2) and so Big-Rip singularity is bypassed. For ${\alpha} \le 3.2,$ the phantom divide line is never crossed, while for ${\alpha} > 4$, the first crossing occurs in the past but second one always occurs in the future. The model fits perfectly with SNIa data for a wide range of values of ${\alpha}$. One can also observe that the state parameter $w_{e}$ remains nearly zero till $z = 2.5$ (fig-2), which confirms that the growth of perturbation in the present model tracks the concordance model closely.\\

\noindent
Thus the present model fits perfectly with the $\Lambda$CDM model without any problem what-so-ever. However, the problem arises while one tries to fit the recently released 7-year WMAP \cite{w1}, \cite{w2} constraint on the redshift of matter-radiation equality at early ISW effect. The value of the integral in equation (16) is $X = \exp{(-\int_t^{t_0}\Gamma dt'}) = 0.3236$, taking ${\alpha} = 4$. As a result the redshift at matter-radiation
equality is pushed far away to $z_{eq} = 4380$, taking only baryons into account, ie., $\Omega_m = \Omega_B = 0.042$. It goes even further if some amount of $\Omega_{CDM}$ is incorporated. The situation is even worse for ${\alpha} > 4$. The whole situation taking some lower values of $\alpha$ and with the same values of $h = 0.7$, $H_0 t_0 = 1$, is depicted in table - 1.\\
\pagebreak
\begin{center} \textbf{Table - 1}\end{center}
\begin{tabular}{|c|c|c|c|c|c|c|c|c|c|}
\hline
${\alpha}$ & $X$ & Nature of $\Gamma$& $z_a$ & $w_{e0}$& $\Omega_m$  & $\Omega_{CDM}$ &$w_{cm0}$ & $z_{eq}$ & Fit with SNIa \\
  \hline
  4   & 0.3236 & Rises from $z = 30$   & 0.56 & -1    & 0.04 & Nil  & -1.20 & 4380 & Indistinguishable from $\Lambda$CDM.\\\hline
  2   & 0.1876 & Rises from $z = 2500$ & 1.08 & -0.56 & 0.09 & 0.05 & -0.62 & 3257 & Indistinguishable from $\Lambda$CDM.\\\hline
  1.8 & 0.1563 & Rises from $z = 7000$ & 1.2  & -0.51 & 0.12 & 0.08 & -0.58 & 3217 & Indistinguishable from $\Lambda$CDM.\\\hline
  1.5 & 0.0962 & Very large initially  & 1.46 & -0.44 & 0.18 & 0.14 & -0.545& 3234 & Fit is not the very best.           \\\hline
\end{tabular}
\\

\begin{figure}
[ptb]
\begin{center}
\includegraphics[
height=2.034in, width=3.3797in] {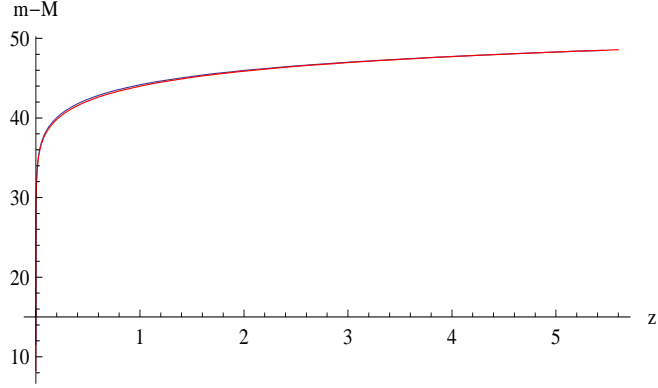} \caption{Distance modulus $(M-m)$ versus redshift $z$ plot of the present model (blue), shows perfect fit with the $\Lambda$CDM model (red) for ${\alpha} =4$. In fact, for
$ {\alpha}> 1.5$ the fit is perfect and it remains so, upto ${\alpha} = 200$.}
\end{center}
\end{figure}

\begin{figure}
[ptb]
\begin{center}
\includegraphics[
height=2.1in, width=3.37in] {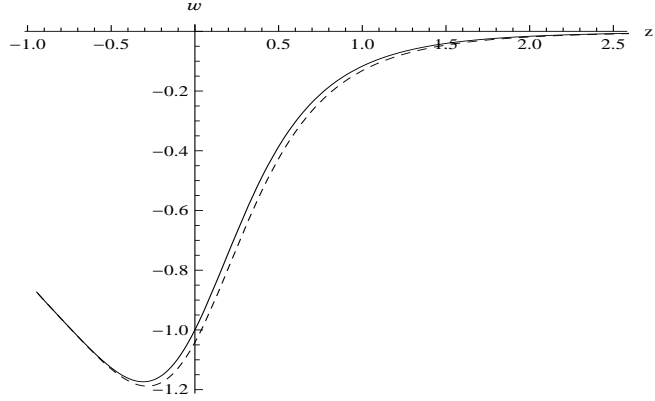} \caption{State parameters $w_{e}(z)$ (continuous) and $w_{cm}(z)$ (dashed) have been plotted against the red-shift parameter $z$ for ${\alpha} = 4$. Both remain nearly zero till $z = 2.5$. The transition redshift is $z_a\approx 0.56$. The present value of $w_e$ is $-1$. A smooth transient crossing for $w_{cm}$ is observed. The first crossing occurs in the past while the other will occur in the future. For ${\alpha} < 3.2$, no such crossing is observed.}
\end{center}
\end{figure}

\begin{figure}
[ptb]
\begin{center}
\includegraphics[
height=2.034in, width=2.8in] {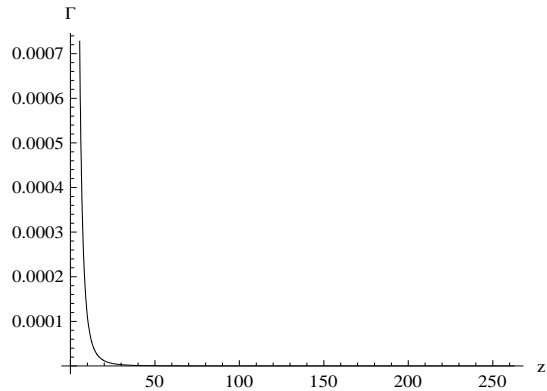} \caption{Creation rate $\Gamma$ versus red-shift parameter $z$
for ${\alpha} = 4$, shows that it is nearly vanishing in the past and started developing only recently at $z\approx 30$.
For higher values of ${\alpha}$, $\Gamma$ starts developing from even smaller redshift value, while for ${\alpha} < 4$,
it starts earlier. Thus, WMAP constraint on $z_{eq}$ can be fitted for ${\alpha} < 3.5$, accommodating some amount
of CDM. }
\end{center}
\end{figure}

\noindent
We are in search of a model where additional creation starts some time after the matter-radiation equality $z_{eq}$. $\Gamma - z$ plot (figure 3) depicts that the creation rate is almost vanishing $(\sim 10^{-8})$ till $z = 30$ for ${\alpha} = 4$ and only thereafter it rises sharply. However the problem is that the creation rate is not sufficiently large to fit WMAP constraint on $z_{eq}$. For ${\alpha} > 4$, the situation is even worse. The behaviour is the same also for lower values up-to ${\alpha} = 2$. The only difference is that the creation starts earlier in this case at $z = 2500$, and so created matter is a little large. Thus WMAP constraint on $z_{eq}$ is satisfied taking into account only a small amount of CDM, viz., $\Omega_{CDM} = 0.05$. Thus the problem with lss model is encountered here too. For ${\alpha} \le 2$, creation starts very early, much before $z_{eq}$ and so these cases are discarded.\\

\section{Case 2}
The main problem, we repeat, associated with the above model and the lss model \cite{l:s} is that the creation of matter is not sufficient to fit WMAP constraint on the redshift of matter-radiation equality at early ISW effect, accommodating $22\%$ that are constituted by dark matter created in the very early Universe. Thus, the choice of $\Gamma$ should be such that the creation of matter starts near $z \approx 3000$ and in the later epoch it should increase considerably, to make $X$ sufficiently small. So, to find a suitable form of $\Gamma$, we try with a scale factor associated with the so called intermediate inflationary solution \cite{if}, viz.,

\be a = a_0 \exp{[A t^f]},\ee

\noindent $a_0$ being a constant. Such a solution for $A > 0$ and $0 < f < 1$, was presented by Barrow\cite{if}, and was shown to lead to late time acceleration \cite{aks} in different models. To appreciate the underlying beauty of the ansatz (25), let us expand it as,

\[a = a_0[1 + A t^f + \frac{1}{2!}A^2 t^{2f} + .....],\]

\noindent
and observe that for $f = \frac{2}{3}$, the standard matter dominated era of Friedmann model is recovered in the early Universe when second term dominates, and the third term becomes responsible for accelerated expansion in the late stage of cosmological evolution. For $f = \frac{1}{3}$, the third term leads to the standard Friedmann model and acceleration starts a little late. For even smaller values of $f$, the model tracks decelerated expansion for a longer time recovering the standard Friedmann model at some intermediate stage of evolution and leads to accelerated expansion at much later stage of cosmic evolution.

\noindent
The redshift parameter $1 + z = \frac{a(t_0)}{a(t)}$, where, $t_0$ is the present time, is found as

\be 1 + z = \exp[{A(t_0^f - t^f)}]. \ee

\noindent Hence, the Hubble parameter takes the following form,
\be  H = \frac{A f}{t^{(1-f)}} = \frac{A f}{[t_0^f - \frac{\ln(1+z)}{A}]^{\frac{1-f}{f}}} ,\ee

\noindent where, we have used equation (26) to get the second equality in the expression of $H$. In view of equation (6), we can now find a form of $\Gamma$ as,

\be \Gamma = 3H-2(1-f)\left(\frac{H}{A f}\right)^{\frac{1}{1-f}}.\ee

\noindent This form of $\Gamma$ is clearly different from the $\beta-\gamma$ model \cite{l:s}. The most important difference is that the creation rate $\Gamma$  here, starts developing only when the Hubble parameter,

\be H \ge \left[\left(\frac{3}{2(1-f)}\right)^{(1-f)}Af\right]^{\frac{1}{f}},\ee

\noindent since, as already mentioned in the introduction, $\Gamma < 0$ is not allowed by the second law of Thermodynamics (see appendix). The creation pressure and the creation matter density are now found as,

\be {8\pi G} p_{cm} = -\Gamma H = H\left[3H-2(1-f)\left(\frac{H}{A f}\right)^{\frac{1}{1-f}}\right] .\ee

\be 8\pi G \rho_{cm} = 3H^2 -8\pi G \rho_m,\ee

\noindent where,

\be {8\pi G}\rho_m = {8\pi G}\rho_{m0}(1+z)^3 = \frac{\rho_{m0}}{\rho_c}3H_0^2(1+z)^3 = 3H_0^2 \Omega_m (1+z)^3,\ee

\noindent in which, $\rho_c$, $\rho_{m0}$ and $\Omega_m$ are the present values of critical density, matter density and the matter density parameter respectively. We can find the effective state parameter and also the state parameter of the created matter as,

\be w_e =  -\frac{2\dot H + 3H^2}{3H^2} = -1 + \frac{2}{3}\left(\frac{1-f}{A f t^f}\right).\ee

\be w_{cm} = -\frac{2\dot H + 3H^2}{3H^2 -8\pi G \rho_{m}} = - \frac{3A^2f^2 - 2Af(1-f)(t_0^f - \frac{\ln(1+z)}{A})^{-1}}{3A^2f^2 - 3\Omega_{m}H_0^2(1+z)^3[t_0^f -\frac{\ln(1+z)}{A}]^{\frac{2(1-f)}{f}}},\ee

\noindent  Now let us see how far this model parametrized by the two parameters $A$ and $f$ fits with the observed data. We have kept $0.96 \le H_0t_0 \le 1$ and $0.67 \le h (= \frac{9.78}{H_0^{-1}} ~Gyr^{-1}) \le 0.7$, as par with HST project \cite{hst}. As already mentioned, the restriction on the parameters are, $A > 0$ and $0 < f < 1$. We have tested the model by choosing $A$ and $f$ which fit SNIa data, from a wide range of values between $0.08 \le A \le 25$ and $0.03 \le f \le 0.99$. The fit requires large $A$ for small $f$ and vise-versa. We have presented our results briefly in the following table - 2, taking only some integral values of $A$ starting form $A = 15$, since for lower values this model does not probe to large redshift $z$. We have taken $z_{eq} = 3300$, which is very much at par with recently released WMAP data \cite{w1}, \cite{w2} and $\Omega_B = 0.042$, to find the amount of matter produced in the late stage of cosmic evolution restricting the amount CDM produced in the very early Universe.\\

\noindent \textbf{Fitting the observational data.}
\begin{center} \textbf{Table - 2}\end{center}
\begin{center}
\begin{tabular}{|c|c|c|c|c|c|c|c|c|}
\hline
 \;\;$A$ \;\;& \;\;$f$ \;\;& \;\;$z_{\Gamma=0}$\;\; & \;\;$z_{a}$ \;\;& \;\;$w_{e0}$ \;\;&\;\;$\Omega_m$\;\;&\;\;$\Omega_{cm}$\;\;&\;\;$\Omega_{CDM}$\;\; \\\hline

15 &0.056 &468 & 0.70 &-0.35&0.246&0.754&0.204\\\hline
16 &0.053 &658 & 0.71 &-0.35&0.249&0.751&0.207\\\hline
17 &0.051 &1145& 1.32 &-0.36&0.255&0.745&0.213\\\hline
18 &0.048 &1350& 0.82 &-0.35&0.255&0.745&0.213\\\hline
19 &0.046 &2050& 1.04 &-0.36&0.257&0.743&0.215\\\hline
20 &0.044 &2914& 1.09 &-0.36&0.258&0.742&0.216\\\hline
21 &0.0415&3080& 0.40 &-0.34&0.257&0.743&0.215\\\hline
22 &0.0392&3157&-0.11 &-0.33&0.256&0.744&0.214\\\hline
\end{tabular}\\
\end{center}

\noindent
In table 2, $z_{\Gamma=0}$ and $z_a$ symbolize the redshift values at which the creation of matter and the acceleration start respectively, while $w_{e0}$ is the present value of effective state parameter.  Let us list our observation point by point.\\

\noindent
1. The distance modulus versus redshift curve fits between the present and the $\Lambda$CDM model (taking $\Omega_{\Lambda} = 0.74$ and $\Omega_m = 0.26$) almost perfectly for a wide range of values of the parameters $A$ and $f$. \\
2. It is observed that for the combinations of $A$ and $f$, which can probe to a distant redshift, the present value of the state parameter is nowhere near $-1$, yet, the model fits both the experimental data, viz., SNIa and WMAP. Particularly, for $A \ge 22$, the acceleration is yet to start.\\
3. The most important point is to note that for $A > 10$, $z_{eq}$ is at par with the recent 7-year WMAP data \cite{w1}, \cite{w2}, only if $24\% - 26\%$ of matter (baryons and CDM) is assumed to have formed in the very early Universe. The table shows that the density parameter $0.74 \le \Omega_{cm} \le 0.76$, which corresponds to $74\% - 76\%$ of matter created in the matter dominated era. Thus, instead of taking into account $74\%$ of dark energy, creation of dark matter by the same amount in the matter dominated era, solves the cosmic puzzle.\\
4. The behaviour of the creation parameter $\Gamma$ given in equation (28) has been plotted in figure (4) for a particular pair of the parameters $A = 19$ and $f = 0.046$. It shows that the creation started at $z = 2050$, reaches a maxima during reionization era and presently it is insignificantly small. The behaviour is the same for all other pairs of $A$ and $f$, only the redshift values at which creation starts ($z{_\Gamma=0} $) and its maxima changes.\\
5. Taking the same values of $A$ and $f$, figure (5) has been plotted. It represents the combined plot of effective state parameter $w_e$ and the state parameter $w_{cm}$ corresponding to created matter, versus redshift parameter $z$, since creation started. These two figures (4) and (5) depict that though creation started rather early at $z = 2050$ and reaches its maxima around $z = 1100$, acceleration started only recently at $z = 1.04$. Using relation (26) it is found that it requires nearly $13.99~ Gyr$ to create $74\%$ of matter. On the other hand, inflation is supposed to start at $10^{-42}s.$, and ends at around $10^{-32} s$. Thus $22\%$ that are constituted by dark matter has been created in $10^{-32}s$ only, in the very early Universe. This gives a comparison of creation phenomena in curvature dominated and low curvature regions.\\
6. We have also presented a suitable contour plot in figure (6), to explore the data presented in table-2 at a glance. The plot presents all the successful combinations of the parameters $A$ and $f$, which fit SNIa data, and satisfy WMAP constraint on $z_{eq} = 3300$, keeping $H_0t_0 \approx 1$, and $24\%\le \Omega_{m}\le 26\%$. Calculation shows that WMAP constraint on $z_{eq}$ is not satisfied for lower or higher values of $\Omega_m$, with the same parametric combination of $A$ and $f$. Particularly, for $\Omega_m = 0.2$, $2400\le z_{eq}\le 2500$, while for $\Omega_m = 0.3$, $3800\le z_{eq}\le 3900$. Thus, nearly $26\%$ of primeval matter in the form of baryons and CDM, is required to fit presently observable data, in view of particle creation phenomena.\\

\noindent
Finally, it is no less important to understand if adiabatic process occurs instead, due to the fact that large time taken to (eg., about $14$ Gyr.) particle creation phenomena in the low curvature region. It is known \cite{bd} that if the expansion rate is very weak, the production of high mass particles is exponentially small. This is due to the fact that large amount of energy must emerge from changing gravitational field to supply particle's rest mass. Thus particle number remains adiabatic invariant and a comoving particle detector remains unexcited, which means the probability of detecting particles falls sharply to zero. However, in discussing quantum particle production in curved space time \cite{bd}, the balance law (5) has never been accounted for, which is crucial in the present analysis. The balance law introduces a back reaction phenomena. As soon as some particles are produced, they impart negative pressure $p_{cm}$, which enhances the expansion rate causing more particles of higher mass to produce. The process continues as long as the Universe expands sufficiently so that the curvature fluctuation is further reduced and the creation rate falls. This feature is present in the $\Gamma - z$ plot of figure (4). Thus, it appears that adiabatic process will not be simulated.\\

\noindent

\begin{figure}
[ptb]
\begin{center}
\includegraphics[
height=2.1in, width=3.3in] {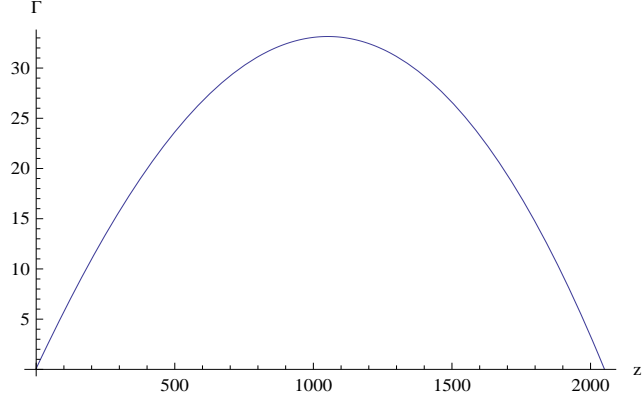} \caption{ The behaviour of $\Gamma$ versus $z$ has been depicted for $A = 19$ and $f = 0.046$. Creation starts in the matter dominated era around $z = 2050$ and its rate has a maxima around $z_{\Gamma_{max}} = 1100 (= z_{recombination})$. Presently the creation rate is insignificantly small. The behavior is the same for all other combinations of $A$ and $f$, only $z_{\Gamma = 0}$ and $z_{\Gamma_{max}}$ are different.}
\end{center}
\end{figure}

\begin{figure}
[ptb]
\begin{center}
\includegraphics[
height=2.1in, width=3.3in] {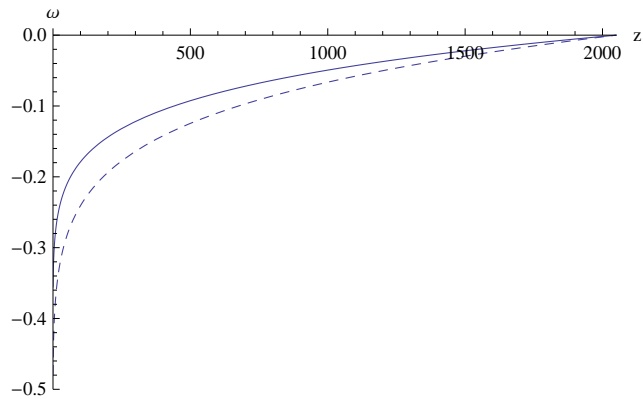} \caption{ The figure shows the combined plot of effective state parameter $w_e$ and the state parameter $w_{cm}$ of created matter, versus redshift parameter $z$, since creation started, taking $A=19$ and $0.046$. While figure-4 depicts that creation started at $z=2050$ and reaches its maxima at $z=1100$, figure-5 shows that most of the time universe undergoes decelerated expansion while acceleration started recently at $z=1.04$.}
\end{center}
\end{figure}

\begin{figure}
[ptb]
\begin{center}
\includegraphics[
height=2.1in, width=3.3in] {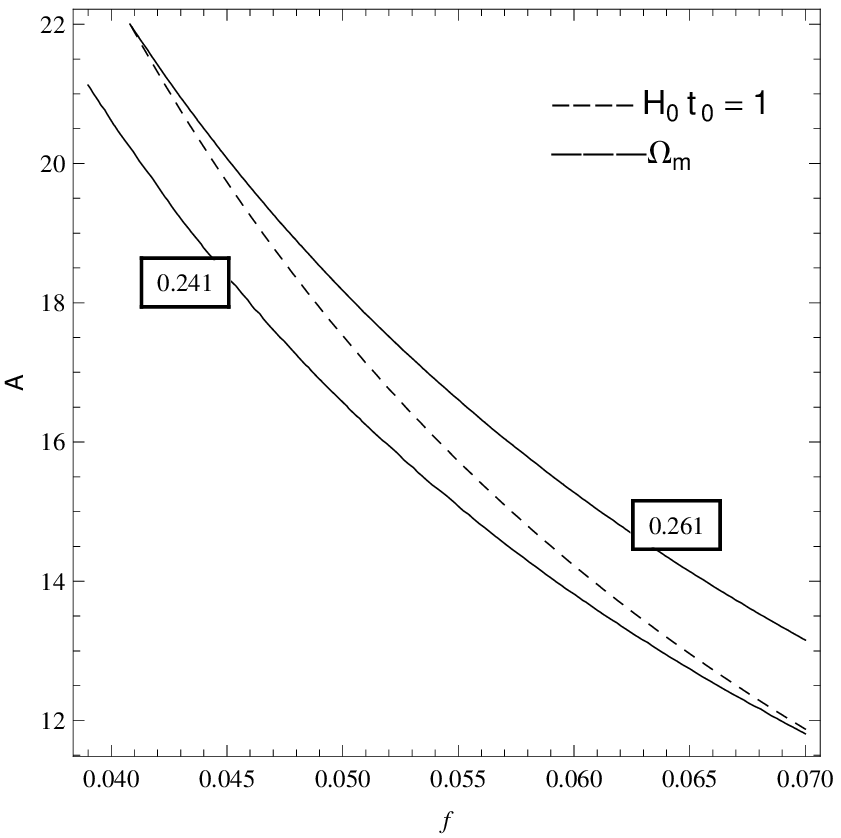} \caption{ The contour plots of $H_0 t_0 = 1$ (dotted line) and $\Omega_m$ for different parametric values of $A$ and $f$ which fit SNIa data, have been combined together. The plot shows that $H_0 t_0 = 1$ line lies within the two lines $\Omega_m = 0.24$ and $\Omega_m = 0.26$, which are calculated taking $z_{eq} = 3300$. Thus, the present model fits SNIa data, satisfies WMAP constraint on $z_{eq}$, demanding the correct amount of $\Omega_m$ required for structure formation and agrees with experimental constraint on $H_0t_0 = 1$.}
\end{center}
\end{figure}

\section{\bf{Concluding remarks}}
We have presented a phenomenological cosmological model based on particle creation in the matter-dominated era, which  fits SNIa data and the redshift of the early integrated Sachs-Wolfe effect at the matter-radiation equality, $z_{eq} = 3145^{+140}_{-139}$ \cite{w1} and $z_{eq} = 3196^{+134}_{-133}$ \cite{w2}, determined by WMAP. The value of h-parameter $(h\approx 0.70)$ and $H_0 t_0 \approx 1$ also are very much within observational limit obtained from HST project \cite{hst}. Such a model constraints the amount of primeval matter to $0.24\le\Omega_m\le0.26$, out of which CDM amounts to $0.20 \le \Omega_{CDM}\le 0.22$, which is again the same amount required for structure formation.\\

\noindent
Quantum particle production phenomena (QFT in CST) has been discussed in the literature in detail \cite{bd}, \cite{mw}. The energy of these particles may then be extracted from the gravitational field \cite{1}. To study the classical consequence of particle creation phenomena, kinetic collision theory may be adopted to find a balance law in addition to standard Einstein's equations, if weakly interacting massive particles (WIMP) are taken into account. In view of such a balance law a cosmological model of the early Universe has been explored \cite{2}. In that model, as the particle production rate becomes comparable to the expansion scalar ($\Theta = 3H$), it builds up a large negative creation pressure that pushes the Hubble parameter $H$ to approximately constant value. As a result, inflationary behaviour due to a large particle production rate is realized. Consequently, the universe starts with a de Sitter phase avoiding cosmological singularity. As inflation continues, the expansion rate becomes too large comparable to the particle production rate. In such a dilute cosmic fluid, particle production rate decreases and halts at some time. At this stage, in the absence of sufficient negative pressure, inflation ends giving way to reheating and the resulting Universe smoothly approaches the familiar Friedmann-Lemaitre-Robertson-Walker behaviour. Since the radiation dominated era, in such a model is of standard Friedmann type (i.e., $a \propto t^{1/2}$, $a$ being the scale factor), so the standard Big-Bang-Nucleosynthesis (BBN) remains unaltered. Thus cosmological evolution of the early Universe may be explained successfully in view of particle creation phenomena.\\

\noindent
Inflation makes the Universe almost spatially flat, which has also been confirmed by recent observations. However, at the end of inflation, if the Universe remains slightly away from spatial flatness, it may cause particle production in the matter dominated era again. If this happens, then even a slow particle production rate may cause sufficient negative pressure in billions of years to cause recently observed cosmic acceleration. This phenomena has been studied earlier \cite{j}, \cite{l:s}, but the model suffered from a clear conflict \cite{g:r} between the low (SNIa) and high redshift (WMAP) data. This problem has been resolved in the present model, which constraints primeval matter to $\Omega_m \approx 26\%$. Since, particle production starts long after matter-radiation equality, so the early growth of perturbation in connection with the structure formation also remains unaltered following the $\Lambda$CDM concordance model closely. Thus particle production process may successfully explain the late stage of cosmic evolution also. As a result, the cosmological evolution of the Universe may be explained successfully in view of particle creation phenomena from early time till date.\\

\noindent
Presently we are having numerous dark energy models, explaining the late time cosmic phenomena. Most of these models do not explain the early Universe at one hand and it is not possible to identify these models from one another in any of the future experiments, on the other. Cosmological consequence of particle creation phenomena does not require dark energy at all at any stage, and some programmes have been taken in the recent years to detect lightest neutralino of roughly $10-10000$ GeV - one of the weakly interacting massive particles (WIMP). Thus, if creation phenomena of
cold-dark-matter can solve the presently observed cosmic puzzle single handedly, without taking into account the dark energy at all, may be resolved in near future.\\

\section{Appendix: Thermodynamics of adiabatic particle creation}

In this appendix we formulate the balance equation in connection with particle creation phenomena. This has been done in \cite{l:s}, \cite{2}. However, the approaches are slightly different, hence we produce a straight forward calculation. Adiabatic cosmological evolution in the presence of particle creation can be treated in the open system, and so the first law of Thermodynamics is modified as,

\be d(\rho V) + p_m ~d V - \frac{h}{n}d(nV) = 0,\ee

\noindent
where, $\rho$, $p_m$, $V$, $n$  and $h$ are the total energy density, the true thermodynamical pressure, any arbitrary co-moving volume, the number of particles per unit volume and the enthalpy per unit volume respectively. In the case under consideration, the system receives heat only due to the transfer of energy from gravitation to matter. So, creation of particles acts as a source of internal energy. Thus for adiabatic transformation the second law of Thermodynamics reads,

\be TdS =  d(\rho V) + p_m d V - \mu d(nV),\ee

\noindent
Combination of the two laws (35) and (36) gives,

\be TdS = \frac{h}{n}d(nV)-\mu d(nV) = T\sigma d N,\ee

\noindent
where, we have used the usual expression for the chemical potential as $\mu n = h - Ts$ and define $s = \frac{S}{V}$ to be the entropy per unit volume and $\sigma = \frac{S}{N}$ as the specific entropy. Thus we observe that the second law of thermodynamics viz., $dS \ge 0$ implies $dN \ge 0$, and the reverse process is thermodynamically impossible, ie., particle can only be created and can not be destroyed. Further, expressing $S$ in terms of $\sigma$, the above equation can also be expressed as,

\be TN d\sigma = 0\Rightarrow \dot\sigma = 0, \ee

\noindent
Hence, in the adiabatic particle creation phenomena, entropy increases, while the specific entropy remains constant. First law given by equation (35) can also be expressed as,

\be Vd\rho+\rho dV+p_m dV-hdV -\frac{hV}{n}dn = 0 \Rightarrow Vd\rho-\frac{hV}{n}dn = 0 \Rightarrow \dot\rho = h\frac{\dot n}{n},\ee

\noindent
Now, the energy-momentum tensor $T^{\mu\nu}$ along with the conservation law when creation phenomena is incorporated are,

\be T^{\mu\nu} = (\rho+p_m+p_{cm})u^{\mu}u^{\nu} - (p_m+p_{cm}) g^{\mu\nu}, ~~~T^\mu _{\nu;\mu} = 0,\ee

\noindent
where, $\rho = \rho_m + \rho_{cm}$ is the total energy density and $p_m$ is the thermodynamic pressure, as already stated, while, $p_{cm}$ is the creation pressure and $u^\mu$ is the component of four velocity vector. The energy conservation law (40) in homogeneous cosmological models reads,

\be \dot\rho + \Theta (\rho + p_m + p_{cm}) = 0,\ee

\noindent
where, $\Theta = 3H$ is the expansion scalar, $H$ being the Hubble parameter. If we now plug in $\dot\rho$ from equation (39) in the above equation (40), we get,

\be p_{cm} = -\frac{\rho + p_m}{\Theta}\left(\Theta + \frac{\dot n}{n}\right) = -\frac{\rho + p_m}{\Theta}\Gamma,\ee

\noindent
where, $\Gamma = \Theta + \frac{\dot n}{n}$ is the creation rate.\\

\noindent
\textbf{Acknowledgement:} The authors would like to thank the referees and the board member in particular, for pointing out several aspects which improved the quality of the manuscript considerably.

\end{document}